# Security Measures for Black Hole Attack in MANET: An Approach

**Rajib Das**
*Research Scholar,* Department of Computer Science
Assam University, Silchar – 788011, India
rajibdas76@gmail.com

**Dr. Bipul Syam Purkayastha**
*Professor,* Department of Computer Science
Assam University, Silchar – 788011, India
bipul_sh@hotmail.com

**Dr. Prodipto Das**
*Asst. Professor,* Department of Computer Science
Assam University, Silchar – 788011, India
prodiptodas2002@rediffmail.com

**Abstract**

*A Mobile Ad-Hoc Network is a collection of mobile nodes that are dynamically and arbitrarily located in such a manner that the interconnections between nodes are capable of changing on continual basis. Due to security vulnerabilities of the routing protocols, wireless ad-hoc networks are unprotected to attacks of the malicious nodes. One of these attacks is the Black Hole Attack. In this paper, we give an algorithmic approach to focus on analyzing and improving the security of AODV, which is one of the popular routing protocols for MANET. Our aim is on ensuring the security against Black hole attack. The proposed solution is capable of detecting & removing Black hole node(s) in the MANET at the beginning. Also the objective of this paper is to provide a simulation study that illustrates the effects of Black hole attack on network performance.*

1. **INTRODUCTION**

Wireless ad-hoc networks are composed of autonomous nodes that are self- managed without any infrastructure [1]. In this way, ad-hoc networks have a dynamic topology such that nodes can easily join or leave the network at any time. They have many potential applications, especially, in military and rescue areas such as connecting soldiers on the battlefield or establishing a new network in place of a network which collapsed after a disaster like an earthquake. Ad-hoc networks are suitable for areas where it is not possible to set up a fixed infrastructure. Since the nodes communicate with each other without an infrastructure, they provide the connectivity by forwarding packets over themselves. To support this connectivity, nodes use some routing protocols such as AODV (Ad-hoc On-Demand Distance Vector), DSR (Dynamic Source Routing) and DSDV (Destination-Sequenced Distance-Vector). Besides acting as a host, each node also acts as a router to discover a path and forward packets to the correct node in the network. As wireless ad-hoc networks lack an infrastructure, they are exposed to a lot of attacks [2][3]. One of these attacks is the Black Hole attack. In the Black Hole attack, a malicious node absorbs all data packets in itself, similar to a hole which sucks in everything. In this way, all packets in the network are dropped. A malicious node dropping all the traffic in the network makes use of the vulnerabilities of the route discovery packets of the on demand protocols, such as AODV [4]. In route discovery process of AODV protocol, intermediate nodes are responsible to find a fresh path to the destination, sending discovery packets to the neighbour nodes [5]. Malicious nodes do not use this process and instead, they immediately respond to the source node with false information as though it has fresh enough path to the destination. Therefore source node sends its data packets via the malicious node to the destination assuming it is a true path. Black Hole attack may occur due to a malicious node which is deliberately misbehaving, as well as a damaged node interface. In any case, nodes in the network will constantly try to find a route for the destination, which makes the node consume its battery in addition to losing packets.

## 2. ROUTING PROTOCOLS

The primary goal of routing protocols [6] in ad-hoc network is to establish optimal path (min hops) between source and destination with minimum overhead and minimum bandwidth consumption so that packets are delivered in a timely manner. A MANET protocol should function effectively over a wide range of networking context from small ad-hoc group to larger mobile Multihop networks. As fig 1 shows the categorization of these routing protocols.

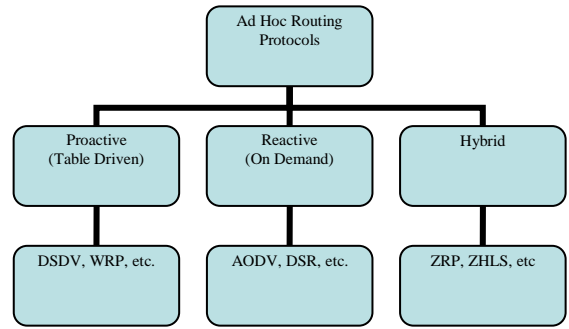

Fig 1: Hierarchy of Routing Protocols

Routing protocols can be divided into **proactive, reactive and hybrid protocols**, depending on the routing topology. Proactive protocols are typically table-driven. Examples of this type include Destination Sequence Distance Vector (DSDV). Reactive or source-initiated on-demand protocols, in contrary, do not periodically update the routing information. It is propagated to the nodes only when necessary. Example of this type includes Dynamic Source Routing (DSR) and Ad Hoc On-Demand Distance Vector (AODV). Hybrid protocols make use of both reactive and proactive approaches. Example of this type includes Zone Routing Protocol (ZRP), ZHLS, etc.

### 2.1. An Overview of AODV Routing Protocol

Ad Hoc On-Demand Vector Routing (AODV) protocol is a reactive routing protocol for ad hoc and mobile networks that maintain routes only between nodes which need to communicate. The AODV routing protocol builds on the DSDV algorithm. AODV is an improvement on DSDV because it typically minimizes the number of required broadcasts by creating routes on an on-demand basis, as opposed to maintaining a complete list of routes as in the DSDV algorithm. The authors of AODV classify it as a pure on-demand route acquisition system, as nodes that are not on a selected path do not maintain routing information. That means, the routing messages do not contain information about the whole route path, but only about the source and the destination. Therefore, routing messages do not have an increasing size. It uses destination sequence numbers to specify how fresh a route is (in relation to another), which is used to grant loop freedom.

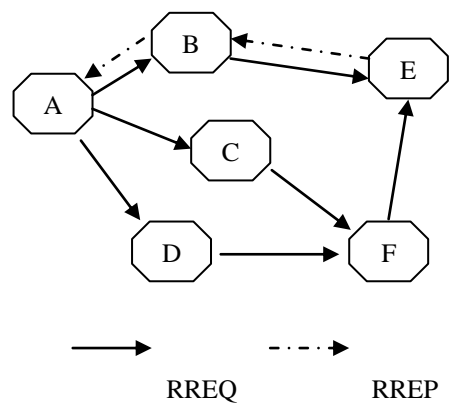

Fig 2: RREQ & RREP message exchange between A & E

Whenever a node needs to send a packet to a destination for which it has no 'fresh enough' route (i.e., a valid route entry for the destination whose associated sequence number is at least as great as the ones contained in any RREQ that the node has received for that destination) it broadcasts a route request (RREQ) message to its neighbors. Each node that receives the broadcast sets up a reverse route towards the originator of the RREQ (unless it has a 'fresher' one).When the intended destination (or an intermediate node that has a 'fresh enough' route to the destination) receives the RREQ, it replies by sending a Route Reply (RREP). It is important to note that the only mutable information in a RREQ and in a RREP is the hop count (which is being monotonically increased at each hop). The RREP travels back to the originator of the RREQ (this time as a unicast). At each intermediate node, a route to the destination is set (again, unless the node has a 'fresher' route than the one specified in the RREP). In the case that the RREQ is replied to by an intermediate node (and if the RREQ had set this option), the intermediate node also sends a RREP to the destination. In this way, it can be granted that the route path is being set up bi-directionally. In the case that a node receives a new route (by a RREQ or by a RREP) and the node already has a route 'as fresh' as the received one, the shortest one will be up dated. The source node starts routing the data packet to the destination node through the neighboring node that first responded with an RREP. The AODV protocol is vulnerable to the well-known black hole attack. This is illustrated in figure 2.

### 3. Black Hole Problem in AODV

Routing protocols are exposed to a variety of attacks. Black hole attack [7] is one such attack and a kind of Denial Of Service (DoS) [8][9] in which a malicious node makes use of the vulnerabilities of the route discovery packets of the routing protocol to advertise itself as having the shortest path to the node whose packets it wants to intercept [10][11]. This attack aims at modifying the routing protocol so that traffic flows through a specific node controlled by the attacker. During the *Route Discovery process*, the source node sends RREQ packets to the intermediate nodes to find fresh path to the intended destination. Malicious nodes respond immediately to the source node as these nodes do not refer the routing table. The source node assumes that the route discovery process is complete, ignores other RREP messages from other nodes and selects the path through the malicious node to route the data packets. The malicious node does this by assigning a high sequence number to the reply packet. The attacker now drops the received messages instead of relaying them as the protocol requires.

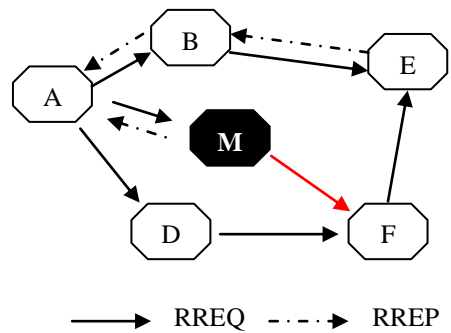

RREQ ----▶ RREP

Fig 3: Black hole Attack in AODV

In the above figure 2, imagine a malicious node 'M'. When node 'A' broadcasts a RREQ packet, nodes 'B' 'D' and 'M' receive it. Node 'M', being a malicious node, does not check up with its routing table for the requested route to node 'E'. Hence, it immediately sends back a RREP packet, claiming a route to the destination. Node 'A' receives the RREP from 'M' ahead of the RREP from 'B' and 'D'. Node 'A' assumes that the route through 'M' is the shortest route and sends any packet to the destination through it. When the node 'A' sends data to 'M', it absorbs all the data and thus behaves like a 'Black hole'.

In AODV, the sequence number is used to determine the freshness of routing information contained in the message from the originating node. When generating RREP message, a destination node compares its current sequence number, and the sequence number in the RREQ packet plus one, and then selects the larger one as RREPs sequence number. Upon receiving a number of RREP, the source node selects the one with greatest sequence number in order to construct a route. But, in the presence of black hole when a source node broadcasts the RREQ message for any destination, the black hole node immediately responds with an RREP message that includes the highest sequence number and this message is perceived as if it is coming from the destination or from a node which has a fresh enough route to the destination. The source assumes that the destination is behind the black hole and discards the other RREP packets coming from the other nodes. The source then starts to send out its packets to the black hole trusting that these packets will reach the destination. Thus the black hole will attract all the packets from the source and instead of forwarding those packets to the destination it will simply discard those. Thus the packets attracted by the black hole node will not reach the destination.

## 4. Proposed Solution

We propose an additional route to the intermediate node that replies the RREQ message to check whether the route from the intermediate node to the destination node exists or not. When the source node receives the FurtherReply (FRp) from the next hop, it extracts the check result from the reply packets. If the result is yes, we establish a route to the destination and begin to send out data packets. If the next hop has no route to the inquired intermediate node, but has a route to the destination node, we discard the reply packets from the inquired intermediate node, and use the new route through the next hop to the destination. At the same time, send out the alarm message to whole network to isolate the malicious node. If the next hop has no route to the requested intermediate node, and it also has no route to the destination node, the source node initiates another routing discovery process, and also sends out an alarm message to isolate the malicious node. Thus we avoid the black hole problem, and also prevent the network from further malicious behavior. But here we assume the black hole nodes do not work as a group and propose a solution to identify a single black hole. However, the proposed method cannot be applied to identifying a cooperative black hole attack involving multiple nodes. We may also develop a methodology to identify multiple black hole nodes cooperating as a group. The technique works with slightly modified AODV protocol and makes use of the Data Routing Information (DRI) table in addition to the cached and current routing tables. A black hole has two properties. First, the node exploits the ad hoc routing protocol, such as AODV, to advertise itself as having a valid route to a destination node, even though the route is spurious, with the intention of intercepting packets. Second, the node consumes the intercepted packets.

### 4.1. Algorithmic approach to avoid black hole attack in MANETs

The solution that we propose here, basically, only modifies the working of the source node without altering intermediate and destination nodes. In this method two main things are added namely Data Routing Information table and cross checking.

**Steps:**
```
1: Source node broadcasts RREQ
2: Source node receives RREP
3: if RREP is from destination or a reliable node
        Then route data packets (source route)
4: Else
     {
            Send further request and identity of intermediate
            Node to next Hop node
        Receive further request, next Hop node of current next hop node, Data Routing Information
        entry for next hope node's next hop. Put a data routing information entry for current
        intermediate node.
5:      if (next hop node is a reliable node)
           {
        Check intermediate node for black hole using data routing information entry
                    if (intermediate node is not a black hole)
                                            route data packets (source route)
                        else
                          {
                                Insecure route
                                Intermediate node is a black hole
                                All the nodes along the reverse path from
        intermediate node to the node that generated
        RREP are black holes (i.e. a malicious node)
                          }
            }
           else
                    Current intermediate node = next hop node
     }
6: Repeat step 4 & 5 until intermediate node is not a reliable node
```

Fig 4: Proposed Algorithm to prevent Black hole Attack

## 4.2. Working Principle

The solution to identify multiple black hole nodes acting in cooperation [12] involves two bits of additional information from the nodes responding to the RREQ of source node S. Each node maintains an additional Data Routing Information (DRI) table.

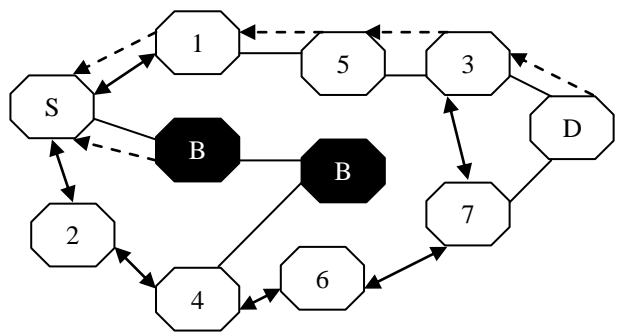

Fig 5: Solution to avoid multiple Black hole attack

In the DRI table, 1 stands for 'true' and 0 for 'false'. The first bit "From" stands for information on routing data packet from the node (in the Node field) while the second bit "Through" stands for information on routing data packet through the node (in the Node field). In reference to the example of Figure 3, a sample of the database maintained by node 4 is shown in Table 1. The entry 1 0 for node 3 implies that node 4 has routed data packets from 3, but has not routed any data packets through 3 (before node 3 moved away from 4). The entry 1 1 for node 6 implies that, node 4 has successfully routed data packets from and through node 6. The entry 0 0 for node B2 implies that, node 4 has NOT routed any data packets from or through B2

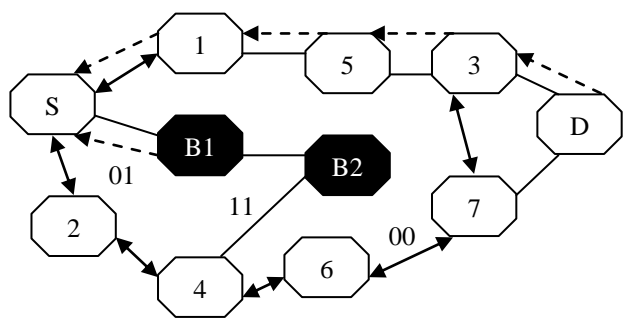

Fig 6: Solution to identify multiple Black hole nodes

We define the following convention for protocol representation in fig 5 & 6.

```
- - - ▶  RREQ/RREP      ——▶  one way propagation
  ◀—▶   Two way propagation   ———  Route indicator
```

*Cross Checking:* In our techniques we rely on reliable nodes (nodes through which the source node has routed data) to transfer data packets. The modified AODV protocol and the algorithm for our proposed methodology are illustrated in Figure 6. In the protocol, the source node (SN) broadcasts a RREQ message to discover a secure route to the destination node. The Intermediate Node (IN) generating the RREP has to provide its Next Hop Node (NHN) and its DRI entry for the NHN. Upon receiving RREP message from IN, the source node will check its own DRI table to see whether IN is a reliable node. If source node has used IN before to route data, then IN is a reliable node and source node starts routing data through IN. Otherwise, IN is unreliable and the source node sends FRq message to NHN to check the identity of the IN, and asks NHN: 1) if IN has routed data packets through NHN, 2) who is the current NHN's next hop to destination, and 3) has the current NHN routed data through its own next hop. The NHN in turn responds with FRp message including 1) DRI entry for IN, 2) the next hop node of current NHN, and 3) the DRI entry for the current NHN's next hop. Based on the FRp message from NHN, source node checks whether NHN is a reliable node or not. If source node has routed data through NHN before, NHN is reliable; otherwise, unreliable. If NHN is reliable, source node will check whether IN is a black hole or not. If the second bit (i.e. IN has routed data through NHN) of the DRI entry from the IN is equal to 1, and the first bit (i.e. NHN has routed data from IN) of the DRI entry from the NHN is equal to 0, IN is a black hole. If IN is not a black-hole and NHN is a reliable node, the route is secure, and source node will update its DRI entry for IN with 01, and starts routing data via IN. If IN is a black-hole, the source node identifies all the nodes along the reverse path from IN to the node that generated the RREP as black hole nodes. Source node ignores any other RREP from the black holes and broadcasts the list of cooperative black holes. If NHN is an unreliable node, source node treats current NHN as IN and sends FRq to the updated IN's next hop node and goes on in a loop from steps 4 through 6 in the algorithm.

## 5. Simulation Environment

We have implemented Black hole attack in an ns-2 [13] simulator. For our simulations, we use CBR (Constant Bit Rate) application, TPC/IP (full duplex communication), IEEE 802.11b MAC and physical channel based on statistical propagation model. The simulated network consists of 30 randomly allocated wireless nodes in a 500 by 500 square meter flat space. The node transmission range is 250-meter power range. Random waypoint model is used for scenarios with node mobility. The selected pause time is 30s seconds. A traffic generator was developed to simulate constant bit rate (CBR) sources. The size of data payload is 512 bytes. In our scenario we take 30 nodes in which nodes 1-22 and 25-30 are simple nodes, and node 23 and 24 are malicious node or Black hole node. The simulation is done using ns-2, to analyze the performance of the network by varying the nodes mobility [14] [15]. The metrics used to evaluate the performance are given below.
**a) Packet Delivery Ratio:** The ratio between the number of packets originated by the "application layer" CBR sources and the number of packets received by the CBR sink at the final destination.
**b) Throughput:** Throughput is the average rate of successful message delivery over a communication channel.
**c) Node Mobility:** Node mobility indicates the mobility speed of nodes.

### 5.1. Result & Discussion

The fig.7 shows the effect to the packet delivery ratio (PDR) measured for the AODV protocol when the node mobility is increased. The result shows both the cases, with the black hole attack and without the black hole attack. It is measured that the packet delivery ratio dramatically decreases when there is a malicious node in the network. For example, the packet delivery ratio is 100% when there is no effect of Black hole attack and when the node is moving at the speed 10 m/s. but due to effect of the Black hole attack the packet delivery ratio decreases to 82 %, because some of the packets are dropped by the black hole node.

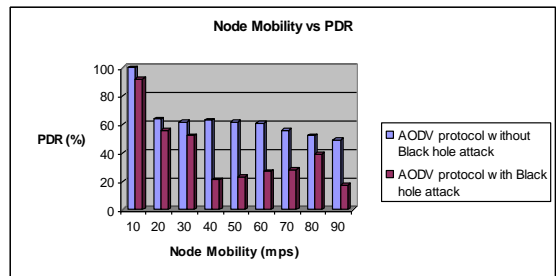

Fig 7: Impact of Black hole attack on PDR

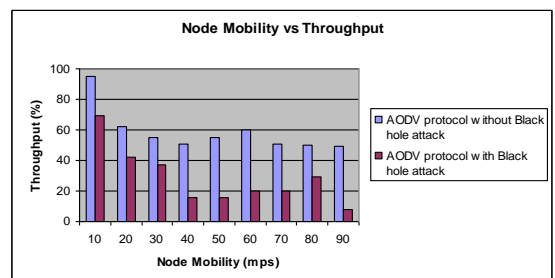

Fig 8: Impact of Black hole attack on Network Throughput

It is observed from the fig.8 that, the impact of the Black hole attack to the Networks throughput. The throughput of the network also decreases due to black hole effect as compared to without the effect of black hole attack. We vary the speed of the node and take the result to the different node speed.

## 6. Conclusion and Future Work

In this paper we have gone through the routing security issues of MANETs, described the black hole attack that can be mounted against a MANET and proposed a feasible solution for it in the AODV protocol. The proposed solution can be applied to a) Identify single and multiple black hole nodes cooperating with each other in a MANET; and b) Discover secure paths from source to destination by avoiding multiple black hole nodes acting in cooperation. Also we showed that the effect of packet delivery ratio and Throughput has been detected with respect to the variable node mobility. There is reduction in Packet Delivery Ratio and Throughput. In Black hole attack all network traffics are redirected to a specific node or from the malicious node causing serious damage to networks and nodes as shown in the result of the simulation. The detection of Black holes in ad hoc networks is still considered to be a challenging task.

We simulated the Black Hole Attack in the Ad-hoc Networks and investigated its affects. In our study, we used the AODV routing protocol. But the other routing protocols could be simulated as well. All routing protocols are expected to present different results. Therefore, the best routing protocol for minimizing the Black Hole Attack may be determined.

**References**

[1] S. Ci *et al.*, "Self-Regulating Network Utilization in Mobile Ad-Hoc Wireless Networks," *IEEE Trans. Vehic. Tech.*, vol. 55, no. 4, July 2006, pp. 1302–10.
[2] M. G. Zapata and N. Asokan, "Securing Ad-Hoc Routing Protocols," *Proc. 2002 ACM Wksp. Wireless Sec.*, Sept. 2002, pp. 1–10.
[3] B. Wu *et al.*, "A Survey of Attacks and Countermeasures in Mobile Ad Hoc Networks," *Wireless/Mobile Network Security*, Springer, vol. 17, 2006.
[4] C. Perkins, E. Belding-Royer, and S. Das, "Ad Hoc On demand Distance Vector (AODV) Routing," IETF RFC 3561, July 2003.
[5] IETF MANET Working Group AODV Draft, http://www.ietf.org/internet-drafts/draft-ietf-manet-aodv-08.txt, Dec 2002.
[6] Elizabeth M. Royer et. al. "A Review of Current Routing Protocols for Ad Hoc Mobile Wireless Networks", IEEE Personal Communication, April 1999.
[7] Dokurer, S.; Erten Y.M., Acar. C.E., SoutheastCon Journal, "Performance analysis of ad-hoc networks under black hole attacks". Proceedings IEEE Volume, Issue, 22-25 March 2007 Page(s):148 – 153.
[8] A. Shevtekar, K. Anantharam, and N. Ansari, "Low Rate TCP Denial-of-Service Attack Detection at Edge Routers," *IEEE Commun. Lett.*, vol. 9, no. 4, Apr. 2005, pp. 363–65.
[9] Mohammad Al-Shurman and Seong-Moo Yoo, Seungjin Park, "Black hole Attack in Mobile Ad Hoc Networks" Proceedings of the 42nd annual Southeast regional conference ACM-SE 42, APRIL 2004, pp. 96-97.
[10] Y-C Hu and A. Perrig, "A Survey of Secure Wireless Ad Hoc Routing," *IEEE Sec. and Privacy*, May–June 2004.
[11] K. Sanzgiri *et al.*, "A Secure Routing Protocol for Ad Hoc Networks," *Proc. 2002 IEEE Int'l. Conf. Network Protocols*, Nov. 2002.
[12] Sanjay Ramaswamy, Huirong Fu, Manohar Sreekantaradhya, John Dixon and Kendall Nygard. **"**Prevention of Cooperative Black Hole Attack in Wireless Ad Hoc Networks". Department of Computer Science, IACC 258 North Dakota State Universities, Fargo, ND 58105.
[13] Network Simulator Official Site for Package Distribution, web reference, http://www.isi.edu/nsnam/ns.
[14] Satoshi Kurosawa, Hidehisa Nakayama, Nei Kato, Abbas Jamalipour, and Yoshiaki Nemoto. "Detecting Blackhole Attack on AODV based Mobile Ad-hoc networks by Dynamic Learning Method". International Journal of Network Security, Vol.5, No.3, PP.338– 346, Nov. 2007.
[15] P. Michiardi, R. Molva. "Simulation-based Analysis of Security Exposures in Mobile Ad Hoc Networks". European Wireless Conference, 2002.

**Acknowledgements**

The authors are grateful to the Head of the Department of Computer Science, Dean, School of Physical Sciences and Hon'ble Vice-Chancellor, Assam University, for providing the facilities to accomplish the present research work.